\documentclass[paper,amsfonts,notoc]{JHEP}

\usepackage{epsfig}

\newcommand{\beq}{\begin{equation}}
\newcommand{\eeq}{\end{equation}}

\newcommand{\bea}{\begin{eqnarray}}
\newcommand{\eea}{\end{eqnarray}}
\newcommand{\nn}{\nonumber}

\def\eqn#1{Eq.~(\ref{#1})}
\def\eqns#1#2{Eqs.~(\ref{#1}) and~(\ref{#2})}
\def\eqnss#1#2{Eqs.~(\ref{#1})-(\ref{#2})}

\def\app#1{Appendix~\ref{#1}}

\newcommand\fverb{\setbox\pippobox=\hbox\bgroup\verb}
\newcommand\fverbdo{\egroup\medskip\noindent%
                        \fbox{\unhbox\pippobox}\ }
\newcommand\fverbit{\egroup\item[\fbox{\unhbox\pippobox}]}
\newbox\pippobox
\def\ord{{\cal O} }
\def\cM{{\cal M}}
\def\eps{\epsilon}
\def\ione{{\bf I}^{(1)}}
\def\itwo{{\bf I}^{(2)}}
\def\htwo{{\bf H}^{(2)}}

\def\MS{$\overline{\rm MS}$}
\def\cg{c_\Gamma}
\newcommand\sss{\scriptscriptstyle}
\newcommand\as{\alpha_{\sss S}} 
\newcommand\gs{g_{\sss S}}
\def\tgs{\tilde\gs}

\title{The high energy limit of QCD at two loops}

\author{Vittorio Del Duca\\ 
Istituto Nazionale di Fisica Nucleare, Sez. di Torino\\
via P. Giuria, 1 - 10125 Torino, Italy\\
        E-mail: \email{delduca@to.infn.it}}

\author{E.~W.~N.~Glover\\
Institute for Particle Physics Phenomenology, 
University of Durham\\ Durham, DH1 3LE, U.K.\\
E-mail: \email{E.W.N.Glover@durham.ac.uk}}

\abstract{ By taking the high-energy limit of the two-loop amplitudes for
parton-parton scattering, we have tested the validity of the loop expansion 
of the high-energy amplitude, arising from a reggeized gluon passed in
the crossed channel.
As expected, we have found that it holds at LL and NLL accuracy,
and hence we have independently re-evaluated
the two-loop Regge trajectory, finding full agreement with the previous
results by Fadin and collaborators.
We have found, though, that the universality implied by the exchange
of a single reggeized gluon in the crossed channel is violated
at the next-to-next-to-leading logarithmic level.}

\keywords{QCD, Jets, NLO and NNLO Computations}
\preprint{{~DCTP/01/66},{~IPPP/01/33},{~DFTT 25/2001},{~hep-ph/0109028}}

\begin{document} 

\section{Introduction}
\label{sec:a}

In the limit of squared center-of-mass energy much greater than
the momentum transfer, $s\gg|t|$, any QCD scattering process is dominated
by gluon exchange in the crossed channel~\footnote{For the sake of
notational simplicity, we omit the carets on the partonic kinematic 
variables.}. Building upon this fact,
the BFKL theory models
strong-interaction processes with two large and disparate scales,
by resumming the radiative corrections to parton-parton
scattering. This is achieved to leading logarithmic (LL) accuracy, in
$\ln(s/|t|)$, through the BFKL 
equation~\cite{Kuraev:1976ge,Kuraev:1977fs,Balitsky:1978ic}, 
i.e. a two-dimensional
integral equation which describes the evolution of the $t$-channel
gluon propagator in transverse momentum space and moment space.
The integral equation is obtained by computing the one-loop LL
corrections to the gluon exchange in the $t$ channel. They are formed
by a real correction: the emission of a gluon along the 
ladder~\cite{Lipatov:1976zz}, and a virtual correction:
the so-called one-loop {\sl Regge\,\, trajectory} (see \eqn{sud}).
The BFKL equation is then obtained by iterating recursively these
one-loop corrections to all orders in $\alpha_s$, to LL accuracy.
The calculation of the building blocks necessary to evaluate the 
next-to-leading logarithmic (NLL) corrections to the BFKL equation
spanned over a decade.
They are the emission of two gluons or two quarks along the 
ladder~\cite{Fadin:1989kf,DelDuca:1996ki,Fadin:1996nw,DelDuca:1996me}, 
the one-loop corrections to the emission of
a gluon along the ladder~\cite{Fadin:1993wh,Fadin:1994fj,Fadin:1996yv,
DelDuca:1999cx,Bern:1998sc}, and the
two-loop Regge trajectory~\cite{Fadin:1995xg,Fadin:1996tb,Fadin:1996km,
Blumlein:1998ib}. The NLL corrections to 
the BFKL equation itself have been 
computed in Refs.~\cite{Fadin:1998py,Camici:1997ij,Ciafaloni:1998gs}. 

In this paper we explicitly take the high-energy limit of the 
two-loop amplitudes
for parton-parton scattering~\cite{Anastasiou:2001sv,Anastasiou:2001kg,
Anastasiou:2001ue,Glover:2001af}. This allows us to
re-evaluate, in a fully independent way, the two-loop
Regge trajectory, and to explore the
possibility of extending the BFKL resummation beyond NLL accuracy.

\section{Virtual corrections in the high-energy limit}
\label{sec:b}

In the high-energy limit $s\gg |t|$, any scattering process is dominated
by gluon exchange in the $t$ channel. In this context, the simplest
process is parton-parton scattering, for which gluon exchange in the $t$ 
channel occurs already at leading order (LO) in perturbative QCD. 
Thus we shall use it
as a paradigm. The amplitude for parton-parton scattering 
$i_a\, j_b\to i_{a'}\,j_{b'}$, with $i, j$ either a
quark or a gluon, may be written as \cite{Kuraev:1976ge}, 
\begin{equation}
\cM^{(0) aa'bb'}_{ij\to ij} = 2  s
\left[\gs\, (T^c_r)_{aa'}\, C^{i(0)}(p_a,p_{a'}) 
\right]
{1\over t} \left[\gs\, (T^c_r)_{bb'}\, C^{j(0)}(p_b,p_{b'}) 
\right]\, ,\label{elas}
\end{equation}
where $a, a', b, b'$ represent the colours of the scattering partons,
and $r$ represents either
the fundamental $(F)$ or the adjoint $(G)$ representations of $SU(N)$, with
$(T^c_{\sss G})_{ab} = i f^{acb}$ and 
${\rm tr}(T^c_{\sss F} T^d_{\sss F}) = \delta^{cd}/2$.
The coefficient functions $C^{i(0)}$, which yield the LO impact factors, 
are given in Ref.~\cite{Kuraev:1976ge} in terms of their spin structure 
and in Ref.~\cite{DelDuca:1995zy,DelDuca:1996km} at fixed
helicities of the external partons. 
The square of the amplitude (\ref{elas}), integrated over the phase
space, yields the parton-parton production rate to LO, $\ord(\as^2)$, in the
high-energy limit.
%For gluon-quark or quark-quark scattering, we only need to exchange
%the structure constants with colour matrices in the fundamental representation
%and change the vertices $C^{g (0)}$ to $C^{q (0)}$ 
%\cite{Kuraev:1976ge,Combridge:1984jn,}.

The virtual radiative corrections to eq.~(\ref{elas}) in
LL approximation are obtained, to all orders
in $\as$, by replacing \cite{Kuraev:1976ge}
\begin{equation}
{1\over t} \to {1\over t} 
\left({s\over -t}\right)^{\alpha(t)}\, ,\label{sud}
\end{equation}
in eq.~(\ref{elas}), where $\alpha(t)$ is related to the one-loop 
transverse-momentum integration. In 
dimensional regularization in $d=4-2\epsilon$ dimensions, it can be written as
\begin{equation}
\alpha(t) = \gs^2\, C_A\, {2\over\epsilon} 
\left(\mu^2\over -t\right)^{\epsilon} c_{\Gamma}\, ,\label{alph}
\end{equation}
with $C_A = N$, and
\begin{equation}
c_{\Gamma} = {1\over (4\pi)^{2-\epsilon}}\, {\Gamma(1+\epsilon)\,
\Gamma^2(1-\epsilon)\over \Gamma(1-2\epsilon)}\, .\label{cgam}
\end{equation}
The fact that higher order corrections to gluon exchange in the $t$ channel can
be accounted for by dressing the gluon propagator with the exponential of
\eqn{sud} is what is called the reggeization, or the Regge trajectory,
of the gluon, and, as said in the Introduction,
lies at the core of the BFKL program.

\begin{figure}[t]
\begin{center}
~\epsfig{file=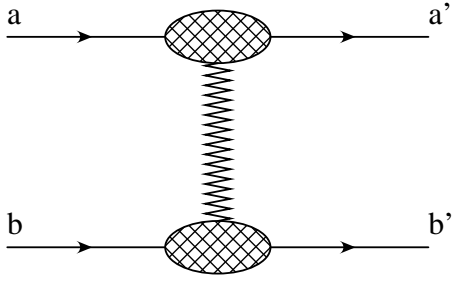,width=6cm}
\end{center}
\caption{The symbolic representation of the factorised form for the high energy
limit of the parton-parton scattering amplitude.  The blobs represent the
coefficient functions $C^{i}(p_a,p_{a'})$ (for $i=g,~q$) while the zigzag line
describes the reggeized gluon exchange. }
\label{fig:full}
\end{figure}

In order to go beyond the LL approximation, we need a prescription that 
allows us to disentangle the virtual corrections to the coefficient functions
in \eqn{elas} from the ones that reggeize the gluon (\ref{sud})
within a loop amplitude. Such a prescription is supplied by 
the general form of the high-energy amplitude for parton-parton 
scattering, arising from a single reggeized gluon passed in the crossed 
channel.
For quark-quark scattering, it can be written as~\cite{Lipatov:1989bs},
\begin{eqnarray}
\lefteqn{\cM^{aa'bb'}_{qq\to qq} } \nonumber\\ 
&=& s
\left[\gs\, (T^c_{\sss F})_{aa'}\, C^{q}(p_a,p_{a'}) \right]
{1\over t} \left[\left({-s\over -t}\right)^{\alpha(t)} +
\left({s\over -t}\right)^{\alpha(t)}  \right]
\left[\gs\, (T^c_{\sss F})_{bb'}\, C^{q}(p_b,p_{b'}) \right]
\nonumber\\ &+& {N^2-4\over N^2} s
\left[\gs\, (T^c_{\sss F})_{aa'}\, C^{q}(p_a,p_{a'}) \right]
{1\over t} \left[\left({-s\over -t}\right)^{\alpha(t)} -
\left({s\over -t}\right)^{\alpha(t)}  \right] 
\left[\gs\, (T^c_{\sss F})_{bb'}\, C^{q}(p_b,p_{b'}) \right]
\nonumber\\  &+& \cdots\, ,\label{elasbq}
\end{eqnarray}
and for quark-gluon and gluon-gluon scattering, it is~\cite{Fadin:1993wh},
\begin{eqnarray}
\lefteqn{\cM^{aa'bb'}_{ig\to ig} } \nonumber\\ 
&=& s
\left[\gs\, (T^c_{r})_{aa'}\, C^{i}(p_a,p_{a'}) \right]
{1\over t} \left[\left({-s\over -t}\right)^{\alpha(t)} +
\left({s\over -t}\right)^{\alpha(t)}  \right]
\left[\gs\, (T^c_{\sss G})_{bb'}\, C^{g}(p_b,p_{b'}) \right]
\nonumber\\  &+& \cdots\, ,\label{elasbg}
\end{eqnarray}
with $r = F (G)$ for $i= q (g)$.
\eqns{elasbq}{elasbg} are symbolically represented in Fig.~\ref{fig:full}.
%\beq
%(T^c_{G_A})_{aa'} = i f^{aca'} \quad (T^c_{G_S})_{aa'} = d^{aca'}
%\quad T^c_{F_A} = T^c_F \quad T^c_{F_S} = \sqrt{N^2-4\over N^2} T^c_F\, .
%\eeq
%
The first (second) line of \eqns{elasbq}{elasbg} corresponds to the exchange 
of a reggeized gluon of negative (positive) signature, belonging to the
antisymmetric (symmetric) representation of $SU(N)$. 
The dots at the end of \eqns{elasbq}{elasbg} account for the (yet unknown) 
exchange of three or more reggeized gluons as well as other colour 
structures that vanish when contracted with the tree amplitude. 

In this paper, we are only interested in terms that survive when projected by
tree-level. In multiplying \eqn{elasbg} by the tree amplitude, the symmetric
part of the reggeized gluon does not contribute, since the colour factor 
of the tree quark-gluon and gluon-gluon amplitudes, contains at least
one structure constant, $f^{bdb'}$, which acts as an $s$-channel 
projector~\cite{Bartels:1993ih},
thus singling out the antisymmetric gluon exchange.
Therefore, the positive signature contribution is only 
present for quark-quark scattering (\ref{elasbq}). 

\begin{figure}[t]
(a)
\begin{center}
~\epsfig{file=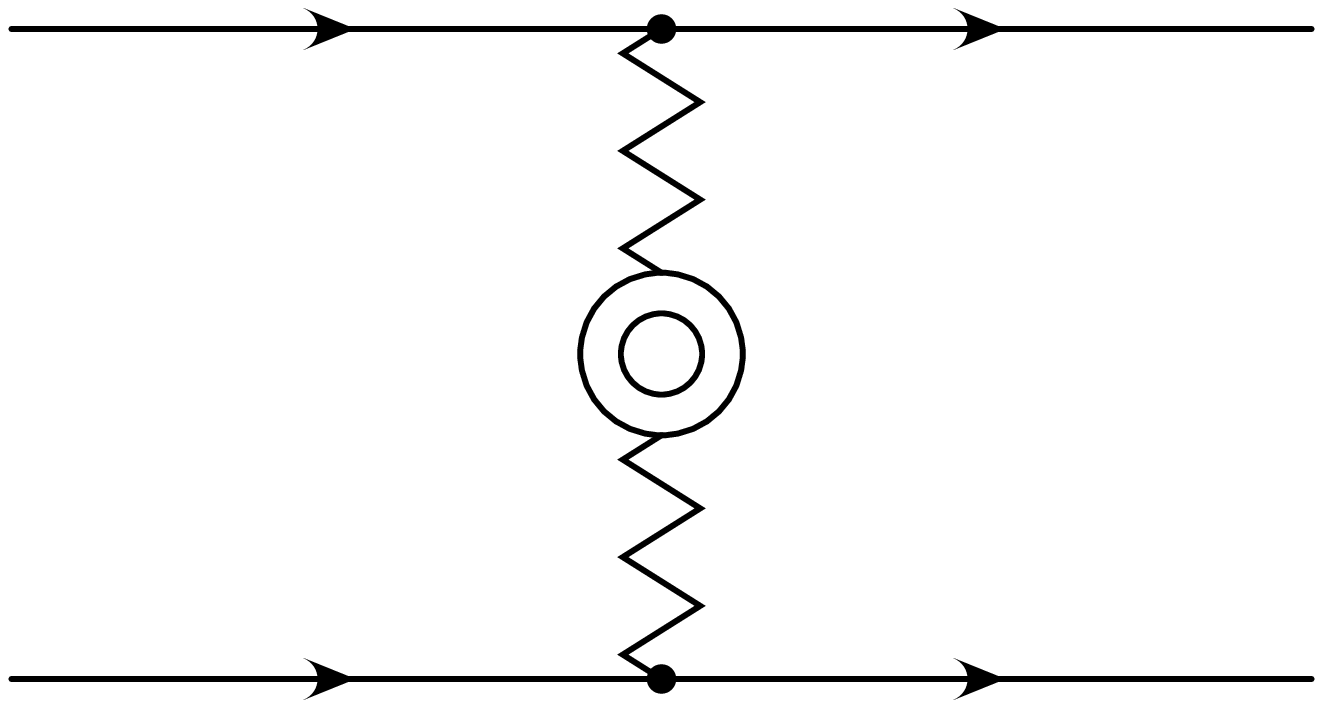,width=4cm}
\end{center}
(b)
\begin{center}
~\epsfig{file=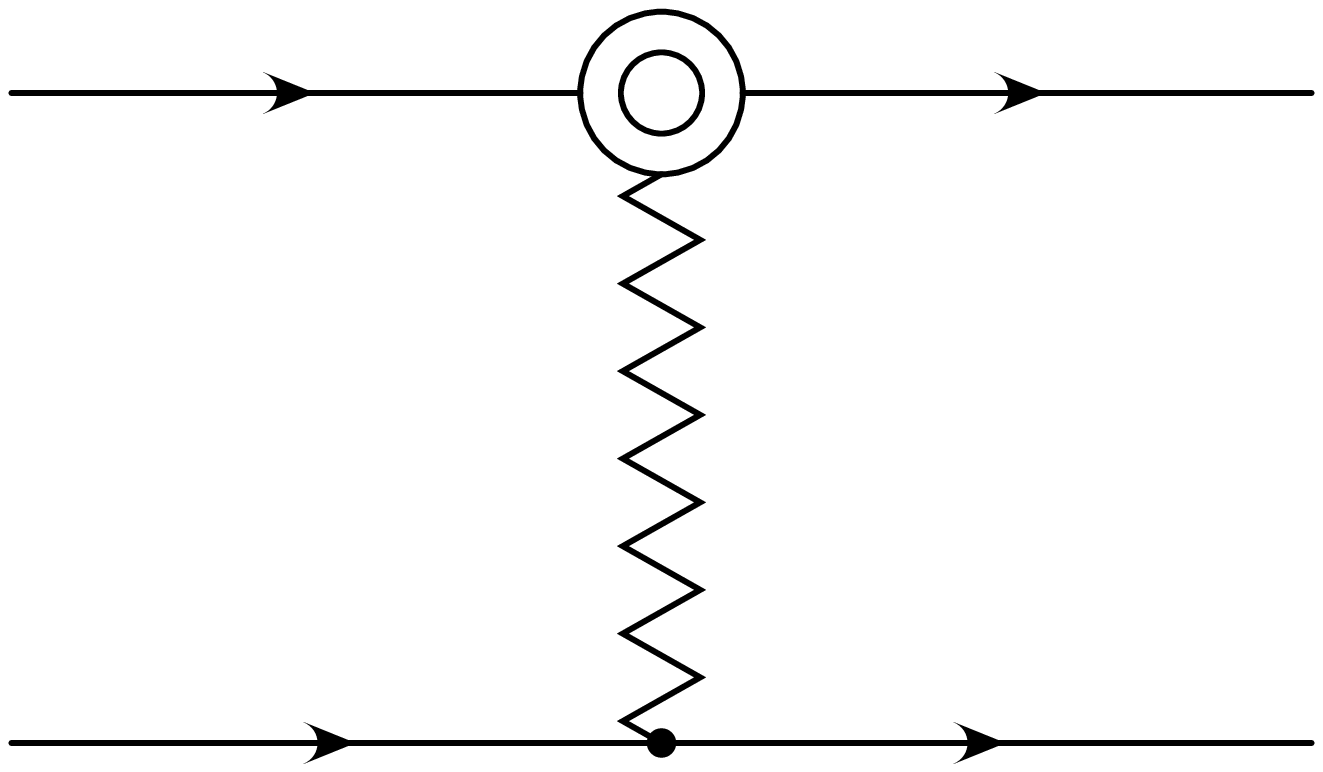,width=4cm}\hspace{2cm} 
~\epsfig{file=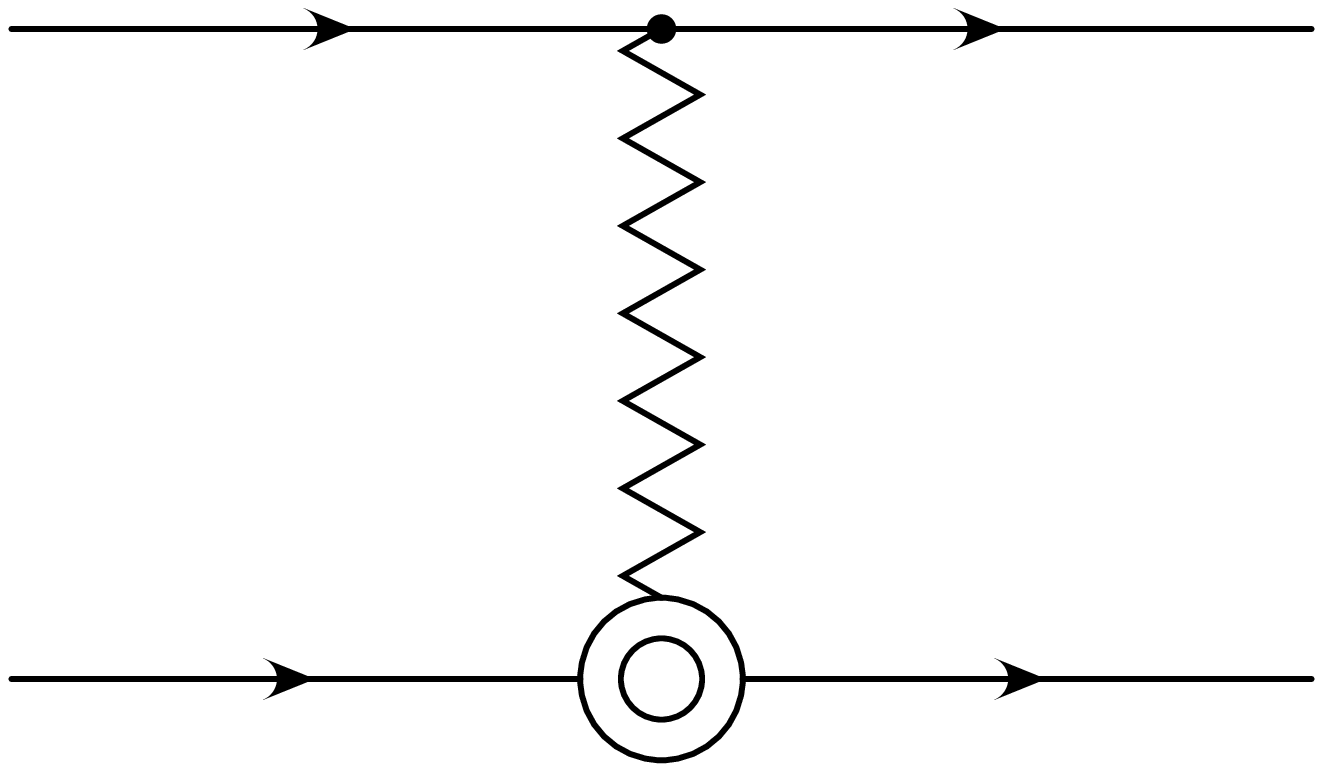,width=4cm} 
\end{center}
\caption{Schematic one-loop expansion of the 
factorised form for the high energy
limit of the parton-parton scattering amplitude. The pairs of concentric circles
represent the one-loop corrections to the impact factor and regge trajectory and
the individual diagrams
represent  terms that contribute at
 (a) leading  and (b) next-to-leading
logarithmic order. }
\label{fig:oneloop}
\end{figure}

The gluon Regge trajectory has the perturbative expansion,
\begin{equation}
\alpha(t) = \tilde\gs^2 \alpha^{(1)}(t) + 
\tgs^4 \alpha^{(2)}(t) + \ord(\tgs^6)\,
,\label{alphb}
\end{equation}
with $\alpha^{(1)}(t)$ given in \eqn{alph}, while the impact factor can be
written as
\begin{equation}
C^{i} = C^{i (0)}(1 + \tgs^2 C^{i (1)} + \tgs^4 C^{i (2)}) + \ord(\tgs^6)\, 
.\label{fullv}
\end{equation}
In \eqns{alphb}{fullv}, we found convenient to rescale the coupling,
\beq
\tgs^2 = \gs^2 \cg \left({\mu^2\over -t}\right)^{\eps}\, .\label{rescal}
\eeq
Then we can write the projection of the amplitudes (\ref{elasbq}) and
(\ref{elasbg}) on the tree
amplitude as an expansion proportional to the tree amplitude squared,
\beq
\cM^{aa'bb'}_{ij\to ij} \cM^{(0) aa'bb'}_{ij\to ij}
= |\cM^{(0) aa'bb'}_{ij\to ij}|^2 \left( 1 +
\tgs^2\ M^{(1) aa'bb'}_{ij\to ij} + \tgs^4 M^{(2) aa'bb'}_{ij\to ij}
+ \ord(\tgs^6) \right)\, ,\label{elasexpand}
\eeq
with $i, j = g, q$. 
The one-loop coefficient of \eqn{elasexpand} is, 
\beq
M^{(1) aa'bb'}_{ij\to ij}
= \alpha^{(1)}(t) \ln\left({s\over -t}\right) +\ C^{i(1)} + C^{j(1)}
- i{\pi\over 2} \left( 1 + \kappa {N^2-4\over N^2} \right)
\alpha^{(1)}(t)\, ,\label{exp1loop}
\eeq
where $\kappa = 1$ for quark-quark scattering, and $\kappa = 0$ in the 
other cases. In \eqn{exp1loop}
we used the usual prescription $\ln(-s) = \ln(s) - i\pi$, for $s > 0$.
Schematically, this is illustrated in Fig.~\ref{fig:oneloop}.
The LL reggeization term, $\alpha^{(1)}(t) = 2C_A/\eps$, 
is independent of the type of
parton undergoing the high-energy scattering process (it is 
{\sl universal}). It is also independent of the infrared (IR) 
regularisation scheme which is used. Conversely, the one-loop coefficient 
functions, $C^{i (1)}$, are process and IR-scheme dependent.
The $C^{i (1)}$'s were computed in conventional dimensional 
regularization (CDR)/
't-Hooft-Veltman (HV) schemes~\footnote{At the amplitude level, the 
difference between the CDR and the HV schemes, which resides in the
number of helicities of the external gluons, is ${\cal O}(\eps)$
\cite{Kunszt:1994sd}. This difference only affects the pole structure at the
squared amplitude level.} in 
Ref.~\cite{Fadin:1993wh,Bern:1998sc,Fadin:1992zt,Lipatov:1997ts,
Fadin:1994qb,DelDuca:1998kx}, and
in the dimensional reduction scheme in Ref.~\cite{Bern:1998sc,DelDuca:1998kx}.
According to \eqns{elasbq}{elasbg}, the coefficient functions $C^{i}$ are real,
the imaginary part of the amplitude being yielded by the trajectory.
In addition to octet exchange,
the explicit calculation of the imaginary part of the one-loop amplitude 
in the high-energy limit \cite{DelDuca:1998kx} 
yields other colour structures.
This fact does not invalidate the NLL program, which is based on the
validity of the antisymmetric part of \eqns{elasbq}{elasbg}, since the 
imaginary part of the one-loop amplitude does not contribute to the 
NLL corrections to the BFKL resummation.

\begin{figure}[t]
(a)
\begin{center}
~\epsfig{file=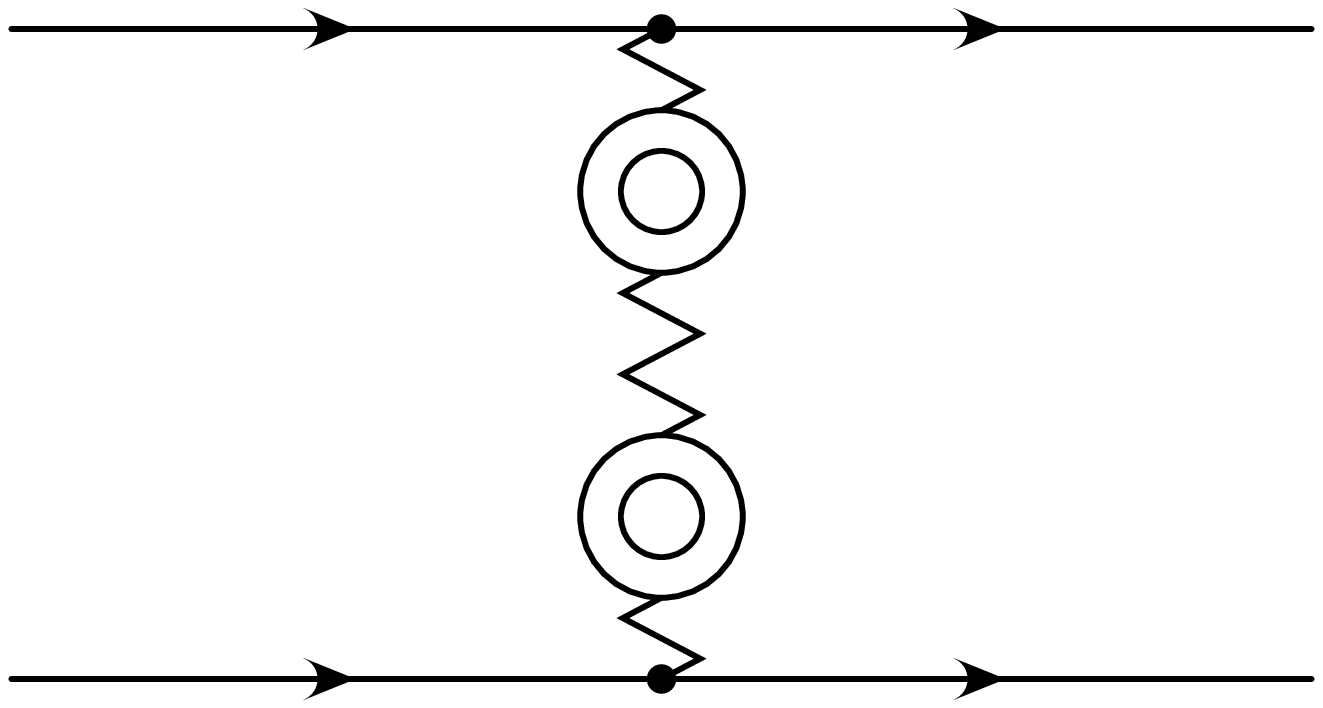,width=4cm}
\end{center}
(b)
\begin{center}
~\epsfig{file=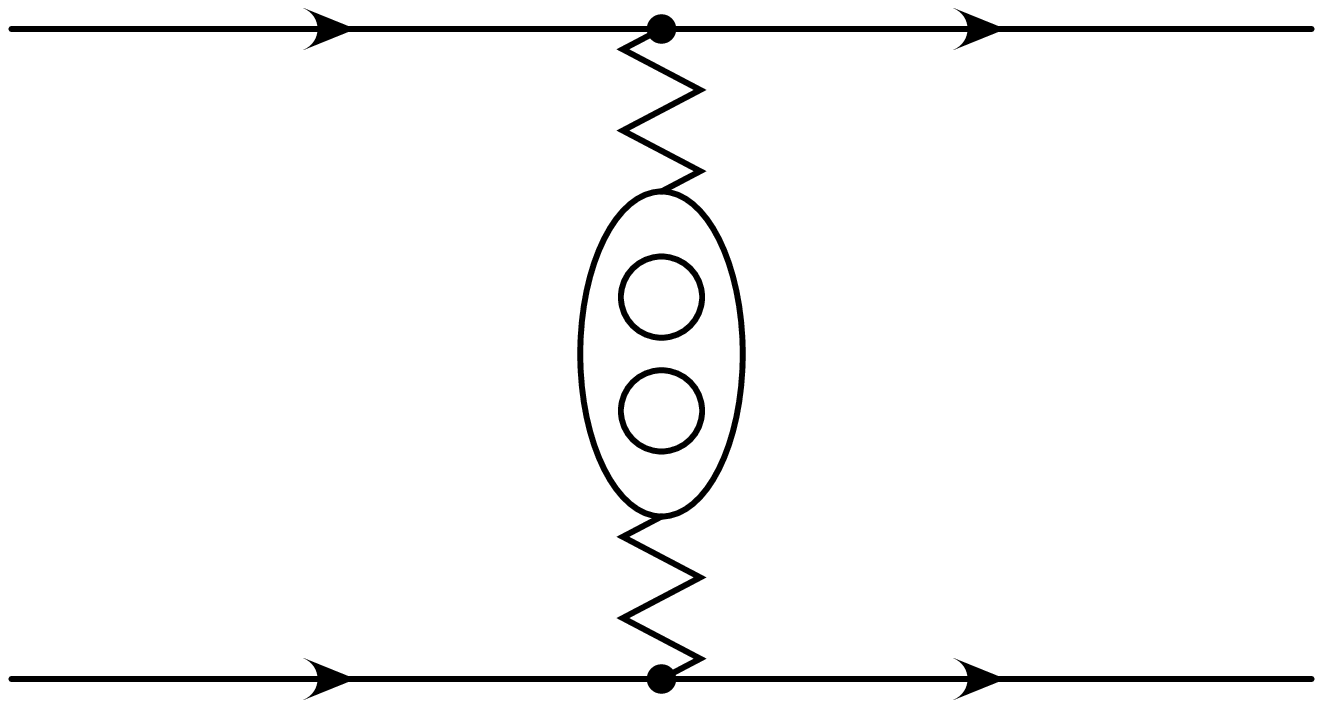,width=4cm}\hspace{1cm} 
~\epsfig{file=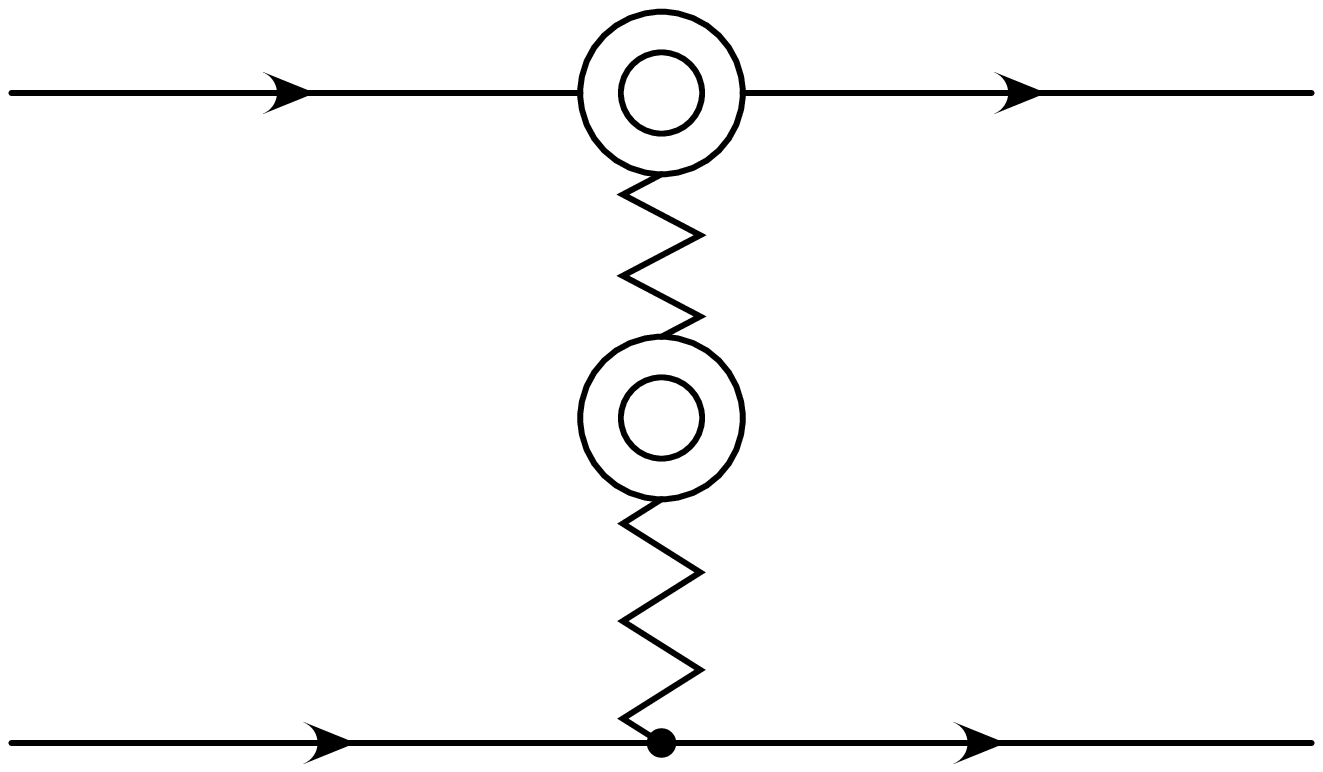,width=4cm} \hspace{1cm} 
~\epsfig{file=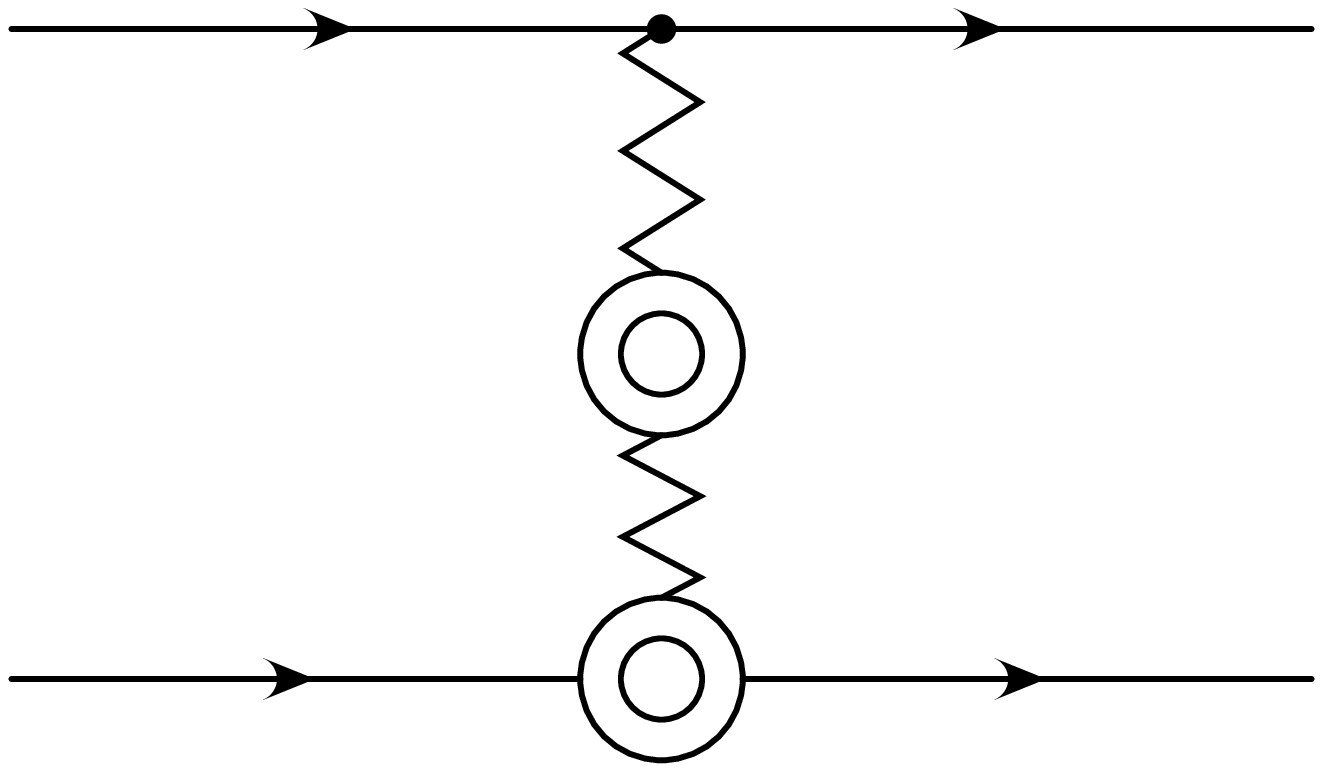,width=4cm} 
\end{center}
(c)
\begin{center}
~\epsfig{file=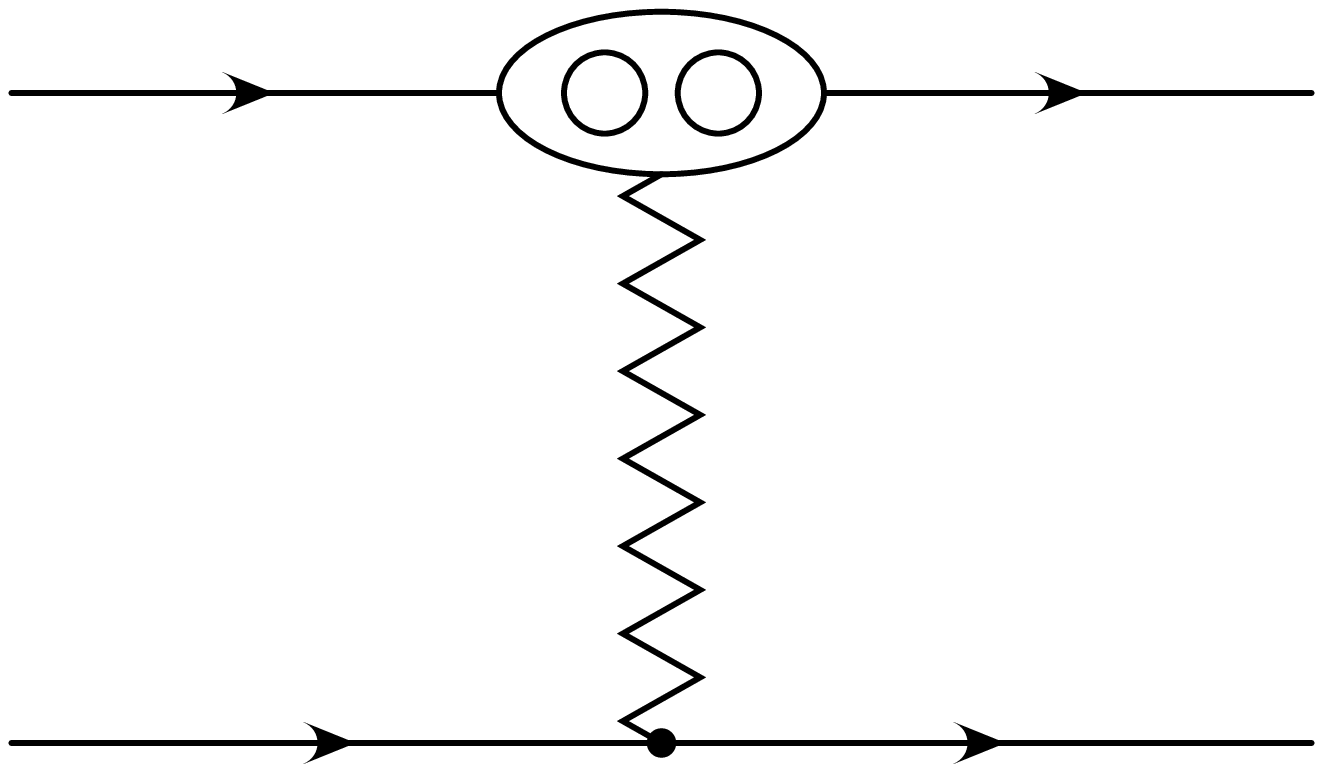,width=4cm}\hspace{1cm}  
~\epsfig{file=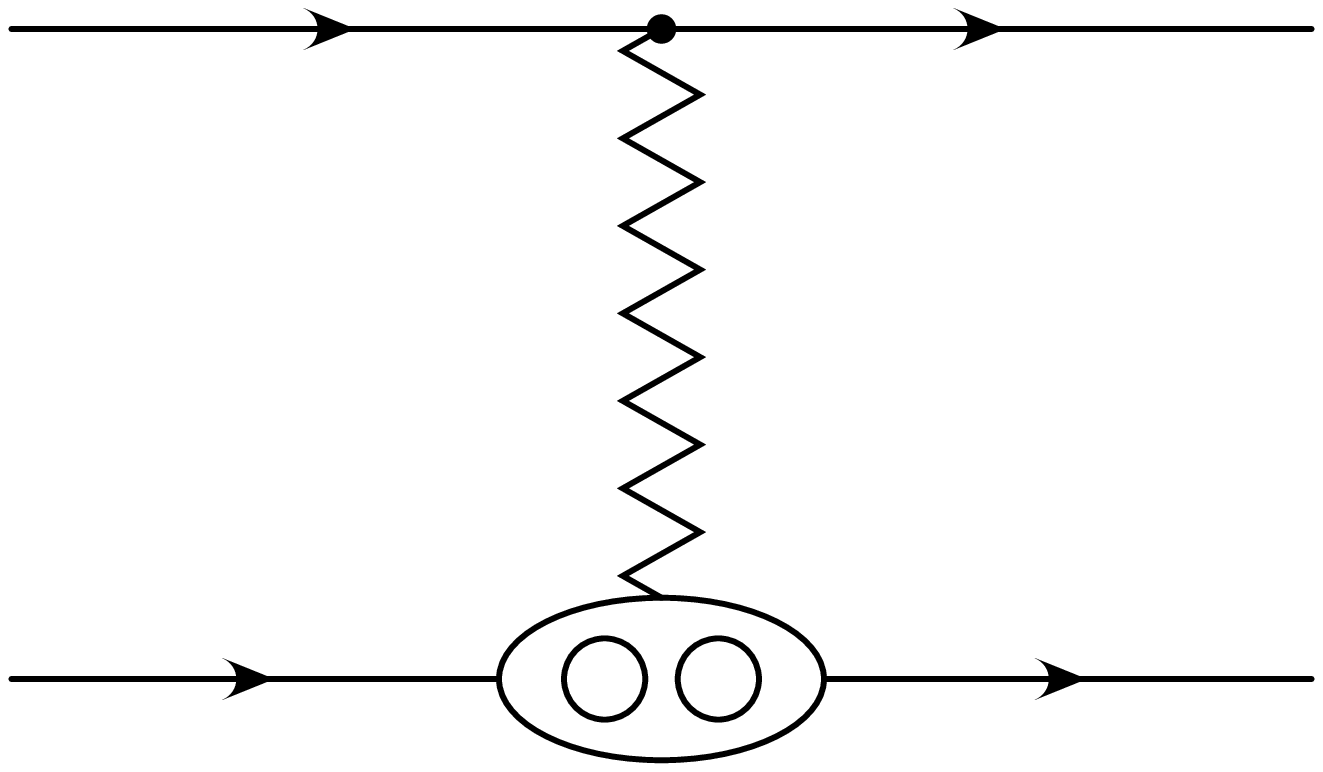,width=4cm}\hspace{1cm}  
~\epsfig{file=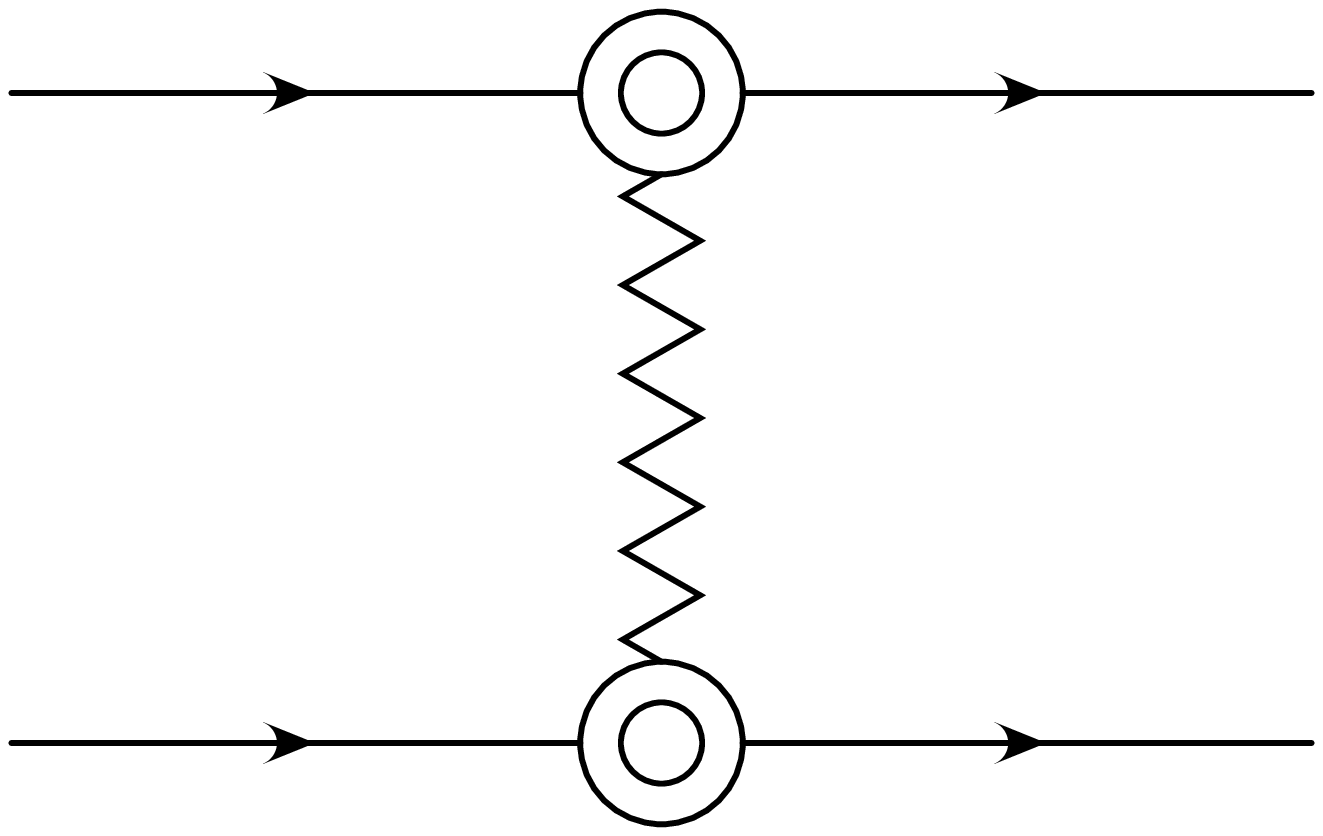,width=4cm} 
\end{center}
\caption{Schematic two-loop expansion of the 
factorised form for the high energy
limit of the parton-parton scattering amplitude. The combinations of ovals and circles
represent the one-loop and two-loop corrections to the impact factor and regge trajectory and
the individual diagrams represent  terms that contribute at
 (a) leading, (b) next-to-leading and (c) next-to-next-to-leading
logarithmic order. }
\label{fig:twoloop}
\end{figure}

The two-loop coefficient of \eqn{elasexpand} is,
\bea
M^{(2) aa'bb'}_{ij\to ij}
&=& {1\over 2} \left(\alpha^{(1)}(t)\right)^2 \ln^2\left({s\over -t}\right) 
\nn\\  &+& \left[ \alpha^{(2)}(t) + \left(C^{i(1)} + C^{j(1)}
\right) \alpha^{(1)}(t) - 
i{\pi\over 2} \left( 1 + \kappa {N^2-4\over N^2} \right)
\left(\alpha^{(1)}(t)\right)^2
\right] \ln\left({s\over -t}\right) \nn\\ &+& 
\left[ C^{i(2)} + C^{j(2)} + 
C^{i(1)}\ C^{j(1)} - 
{\pi^2\over 4} \left( 1 + \kappa {N^2-4\over N^2} \right)
\left(\alpha^{(1)}(t)\right)^2 \right] \nn\\ 
&-& i{\pi\over 2} \left( 1 + \kappa {N^2-4\over N^2} \right)
\left[ \alpha^{(2)}(t) + \left(C^{i(1)} + 
C^{j(1)} \right) \alpha^{(1)}(t) \right]
\, .\label{exp2loop}
\eea
Schematically, this is illustrated in
Fig.~\ref{fig:twoloop}.
The first line of \eqn{exp2loop} is just the exponentiation of the one-loop
trajectory (Fig.~\ref{fig:twoloop}(a)). If the single-log term is known, 
the second line of \eqn{exp2loop}
allows to determine $\alpha^{(2)}(t)$, the two-loop Regge 
trajectory (The first diagram in Fig.~\ref{fig:twoloop}(b)). 
The third and fourth lines are respectively
the real (Fig.~\ref{fig:twoloop}(c)) and the imaginary parts of the constant 
term. Note that the gluon of
positive signature does not contribute to the BFKL resummation at LL and NLL
accuracy. The two-loop
coefficient functions $C^{i(2)}$ could in principle be used to construct
the next-to-next-to-leading
order (NNLO) impact factors, if the BFKL resummation held to 
next-to-next-to-leading-log (NNLL) accuracy.

\section{The two-loop amplitude in the high-energy limit}
\label{sec:c}

As indicated in the Introduction, we wish to make an independent check of the
two-loop trajectory of the reggeized gluon that is
exchanged in parton-parton scattering processes in the high energy limit.
We wish to do this directly by taking the high energy limit of the two-loop parton-parton
scattering amplitudes.   
The interference of the tree- and two-loop amplitudes for each of the
parton-parton scattering processes have
been explicity computed in 
Refs.~\cite{Anastasiou:2001sv,Anastasiou:2001kg,Anastasiou:2001ue,Glover:2001af} 
using conventional dimensional regularization (CDR) and renormalised in the 
\MS\ scheme.   
In these papers the divergent contribution is written in terms of the 
infrared singularity operators $\ione$, $\itwo$ and $\htwo$ proposed by Catani
\cite{Catani:1998bh}
and the tree- and one-loop amplitudes. The
finite remainder is given in terms of logarithms and polylogarithms with
arguments $-u/s$, $-t/s$ and $u/t$.  This latter argument can be flipped using
standard polylogarithm identities so that
the $t \to 0$ limit for the $gg \to gg$, $qg \to qg$ and $qq^\prime \to
qq^\prime$ processes can be
straightforwardly taken.  
After expansion in $\epsilon$, the leading power in $s/t$ of the 
interference between the (unrenormalised) $n$-loop and the tree amplitudes 
for $ij\to ij$ has the form
\begin{equation}
{\rm Re}\ \left(\cM^{(0)*} \cM^{(n)} \right)_{ij\to ij}
= |\cM^{(0)}|^2_{ij\to ij}\ \tgs^{2n}\ 
\sum_{m=0}^n B^{ij}_{nm} \ln^m\left(-{s\over t}\right) \nn
,\label{eq:expand}
\end{equation}
with $\tgs^2$ given in \eqn{rescal}, and where $|\cM^{(0)}|^2_{ij\to ij}$ 
is the high energy limit of the square of the tree-amplitude 
in CDR. For $n=0$, $B^{ij}_{00}=1$.

By explicit comparison of the leading singularity in $t$ with the general
expression given in Eq.~(\ref{eq:expand}), we find the
following relations
\bea
B^{ij}_{11} &=& {2 \over \eps} C_A \label{eq:rela}\\
B^{qg}_{10} &=& {1\over 2} \left( B^{gg}_{10} + B^{qq}_{10}\right) 
\label{eq:relb}\\
B^{ij}_{22} &=& {1\over 2} \left( B^{ij}_{11}\right)^2 \label{eq:relc}\\
B^{ij}_{21} &=&  B^{ij}_{11}B^{ij}_{10} + C_A \beta_0 {2\over \eps^2} + C_A K
{2\over \eps} +C_A^2 \left({404\over 27} - 2\zeta_3\right) 
+ C_AN_F \left(-{56\over 27}\right) \label{eq:reld}
\eea
where
\beq 
\beta_0= {(11C_A-2N_F)\over 6}, \qquad\qquad K = \left({67\over 18} -
{\pi^2\over 6} \right) C_A - {5\over 9} N_F\, .\label{eq:beta}
\eeq
We compare now these relations to \eqns{exp1loop}{exp2loop}.
At this point, a caveat is in order: since we perform the comparison
at the level of the interference with the tree amplitude, we shall miss
any colour structure, which might appear in the two-loop amplitude in
the high-energy limit, but that is projected out by the interference
with the tree amplitude. Therefore a successful comparison
between \eqns{exp1loop}{exp2loop} and \eqnss{eq:rela}{eq:reld} is a
necessary but not sufficient condition for the validity of 
\eqns{elasbq}{elasbg}.

\eqn{eq:rela} verifies the universality of the one-loop trajectory,
and \eqn{eq:relc} its exponentiation at the two-loop level (see the
first line of \eqn{exp2loop})
\beq
B^{ij}_{11} = \alpha^{(1)}(t)\, ;\qquad
B^{ij}_{22} = {1 \over 2} \left(\alpha^{(1)}(t) \right)^2\, .
\eeq
The system formed by \eqn{exp1loop} for
gluon-gluon, quark-quark and quark-gluon scattering is overconstrained,
namely we have three equations and only two unknowns, the one-loop
coefficients $C^{g(1)}$ and $C^{q(1)}$. For instance, we can use the
one-loop amplitudes for gluon-gluon and quark-quark scattering to
determine $C^{g(1)}$ and $C^{q(1)}$, respectively. Then the constant 
term of the amplitude for quark-gluon scattering can be obtained without
any further calculation. Conversely, the explicit calculation of
quark-gluon scattering (see \app{sec:app}) tests \eqn{eq:relb} and thus
the validity of the high-energy expansion to one-loop accuracy,
\bea
C^{g(1)} &=& {1\over 2} B^{gg}_{10}\, ;\quad
C^{q(1)} = {1\over 2} B^{qq}_{10} \, ;\quad
B^{qg}_{10} = C^{g(1)} + C^{q(1)} = {1\over 2} 
\left( B^{gg}_{10} + B^{qq}_{10} \right)\, .
\eea
Comparing \eqn{eq:reld} to the single-log term of \eqn{exp2loop}
determines the value and verifies the universality of the 
two-loop trajectory,
\beq
B^{ij}_{21} = \alpha^{(2)}(t)+\alpha^{(1)}(t)(C^{i(1)}+C^{j(1)})\, 
.\label{singlog}
\eeq
The (unrenormalised) two-loop trajectory is
\bea
\alpha^{(2)}(t) &=& B^{ij}_{21}-B^{ij}_{11}B^{ij}_{10} \nn\\
&=& C_A \beta_0 {2\over \eps^2} + C_A K{2\over \eps} 
+ C_A^2 \left({404\over 27} - 2\zeta_3\right) 
+ C_AN_F \left(-{56\over 27}\right)\, ,\label{eq:2loop}
\eea
in agreement with the unrenormalised two-loop trajectory of
Ref.~\cite{Fadin:1995xg,Fadin:1996tb,Fadin:1996km,
Blumlein:1998ib,lafex}.

If the general form of the high-energy scattering amplitudes (\ref{elasbq}) 
and (\ref{elasbg}) holds to NNLO, and thus the expansion 
(\ref{exp2loop}) is valid up to the constant terms, we can
determine the two-loop coefficient functions $C^{i(2)}$ through
$B^{ij}_{20}$,
\bea
B^{ij}_{20} &=& C^{i(2)} + C^{j(2)} + C^{i(1)}\ C^{j(1)} - 
{\pi^2\over 4} \left(\alpha^{(1)}(t)\right)^2\, 
,\qquad i=q,g\;\; j=g \nonumber\\
B^{qq}_{20} &=& 2 C^{q(2)} + \left(C^{q(1)}\right)^2 - 
{\pi^2\over 4} \left( 1 + {N^2-4\over N^2} \right)
\left(\alpha^{(1)}(t)\right)^2\,
\label{2loopconst}
\eea
As for \eqn{exp1loop}, the system formed by \eqn{2loopconst} for
gluon-gluon, quark-quark and quark-gluon scattering has
three equations and only two unknowns, the one-loop
coefficients $C^{g(2)}$ and $C^{q(2)}$. We can use the two-loop
amplitudes for gluon-gluon and quark-quark scattering to
determine $C^{g(2)}$ and $C^{q(2)}$, respectively. Then 
the validity of the high-energy expansion to two-loop accuracy implies
the relation,
\beq
B^{qg}_{20} - {1\over 4} B^{qq}_{10} B^{gg}_{10} 
- {\pi^2\over 8} {N^2-4\over N^2} \left(\alpha^{(1)}(t)\right)^2
= {1\over 2} \left[
B^{gg}_{20} - {1\over 4} \left( B^{gg}_{10}\right)^2 
+B^{qq}_{20} - {1\over 4} \left( B^{qq}_{10}\right)^2 \right]\, 
.\label{2looprel}
\eeq
Using \eqn{eq:relb}, this can be recast as the difference between
terms depending on the quark-gluon amplitude and terms that depend on
the gluon-gluon and quark-quark amplitudes,
\beq
B^{qg}_{20} - {1\over 2} \left( B^{qg}_{10}\right)^2 -
{1\over 2} \left[
B^{gg}_{20} - {1\over 2} \left( B^{gg}_{10}\right)^2 
+B^{qq}_{20} - {1\over 2} \left( B^{qq}_{10}\right)^2 
\right] -\frac{\pi^2}{2\eps^2} \left(N^2-4\right) = 0
\, .\label{2looprelb}
\eeq
Through the explicit calculation of the $B^{ij}$ coefficients (see
\app{sec:app}) we found that \eqn{2looprelb}
holds for constant and $\zeta_3$ terms but 
that it is violated by terms of ${\cal O}(\pi^2/\eps^2)$,
\bea
\lefteqn{B^{qg}_{20} - {1\over 2} \left( B^{qg}_{10}\right)^2 -
{1\over 2} \left[
B^{gg}_{20} - {1\over 2} \left( B^{gg}_{10}\right)^2 
+B^{qq}_{20} - {1\over 2} \left( B^{qq}_{10}\right)^2 
\right] -\frac{\pi^2}{2\eps^2} \left(N^2-4\right)}\nn \\
&&\hspace{7cm} = \frac{3\pi^2}{\eps^2} \left(\frac{N^2+1}{N^2}\right) + {\cal
O}(\eps)
\, .\label{2loopviol}
\eea
This violation is due to the exchange of three or more reggeized gluons 
which is unaccounted for in \eqns{elasbq}{elasbg}.   

Analogously to \eqn{eq:expand}, we can write 
the leading power in $s/t$ of the interference between the unrenormalised 
$n$-loop and the tree amplitudes as,
\begin{equation}
{\rm Im}\ \left(\cM^{(0)*} \cM^{(n)} \right)_{ij\to ij}
= -{\pi\over 2}\ |\cM^{(0)}|^2_{ij\to ij}\ \tgs^{2n}\ 
\sum_{m=0}^{n-1} D^{ij}_{nm} \ln^m\left(-{s\over t}\right)  
,\label{eq:expandim}
\end{equation}
with $n\ge 1$.

By explicit comparison of the leading singularity in $t$ with the
general expression given in \eqn{eq:expandim}, and by using 
\eqns{alph}{singlog}, we find,
\bea
D^{ij}_{10} &=& {2C_A\over \eps} = \alpha^{(1)}(t) 
\label{eq:irela}\\
D^{ij}_{21} &=&  \left( D^{ij}_{10} \right)^2 = 
 \left(\alpha^{(1)}(t) \right)^2 \label{eq:irelb}\\
D^{ij}_{20} &=&  B^{ij}_{21} =  \left[
\alpha^{(2)}(t)+\alpha^{(1)}(t)(C^{i(1)}+C^{j(1)}) \right]
\label{eq:irelc} 
\eea
for $i=q, g$ and $j=g$, and
\bea
D^{qq}_{10} &=&  {4(4C_F-C_A) \over \eps} \label{eq:ireld}\\
D^{qq}_{21} &=&  {8C_A(4C_F-C_A)\over \eps^2} \label{eq:irele}\\
D^{qq}_{20} &=&  {2(4C_F-C_A)\over C_A}
\left[\alpha^{(2)}(t)+2\alpha^{(1)}(t)C^{q(1)}\right]
\label{eq:irelf}\,  
\eea
for quark-quark scattering. \eqnss{eq:irela}{eq:irelf} are in agreement with 
the imaginary parts of \eqns{exp1loop}{exp2loop}.
However, the same caveat we put forward after \eqn{eq:beta} is valid here,
namely we cannot exclude that other colour structures appear which are killed 
by the projection on the tree amplitude in the high-energy limit.

\section{Conclusions}

By taking the high-energy limit of the two-loop amplitudes for
parton-parton scattering~\cite{Anastasiou:2001sv,Anastasiou:2001kg,
Anastasiou:2001ue,Glover:2001af}, we have
tested the validity of the general form of the high-energy 
amplitudes (\ref{elasbq}) and (\ref{elasbg}) for
parton-parton scattering, arising from a reggeized gluon exchanged in
the crossed channel.
As expected, we have found that it holds at LL and NLL accuracy,
and hence we have independently re-evaluated
the two-loop Regge trajectory (\ref{eq:2loop}), finding full agreement
with Ref.~\cite{Fadin:1995xg,Fadin:1996tb,Fadin:1996km,Blumlein:1998ib}.
We have found, though, that the universality implied by
\eqns{elasbq}{elasbg} is violated at the next-to-next-to-leading order level, 
\eqn{2loopviol}. The source of the discrepancy between 
\eqns{2looprelb}{2loopviol}
might reside in the yet unknown exchange of three reggeized 
gluons, which is unaccounted for in \eqns{elasbq}{elasbg}.

%We have also analysed the imaginary
%part of \eqn{elasb}, and found that while the ansatz (\ref{elasb}) holds
%for gluon-gluon and quark-gluon scattering at two-loop level, it is
%violated already at one-loop level in quark-quark scattering, 
%\eqnss{eq:ireld}{eq:irelf}. This hints that the problem
%with the lack of universality in \eqn{2loopviol} may lie 
%in the quark-quark sector.

Finally, we stress that our comparisons are done at the level of the 
interference between loop and tree amplitudes. 
Other colour structures may be present at the amplitude level, which may be
killed, though, by the projection on the tree amplitude in the high-energy 
limit. Thus a more stringent comparison 
at the amplitude level would be welcome.

\section*{Acknowledgements}

We thank the Les Houches Centre for Physics and the organizers
of the Les Houches Workshop {\em Physics at TeV Colliders} 
where this work was initiated and Ugo Aglietti, Jochen Bartels, Stefano Catani,
Victor Fadin, Lev Lipatov, Alan Martin, Zoltan Trocsanyi and Alan White for helpful 
comments. VDD also thanks the
IPPP for its kind hospitality during the latter stages of this work.

\appendix
\section{The coefficients of the two-loop amplitude in the high-energy limit}
\label{sec:app}
In this appendix we give a complete list of the real and imaginary coefficients $B_{nm}^{ij}$ and 
$D_{nm}^{ij}$ obtained by expanding the interference of tree and 
two-loop graphs
in the high energy limit.  All results are valid in conventional dimensional
regularisation.   The one-loop coefficients are expanded keeping terms through
to
${\cal O}(\epsilon^2)$ while the two-loop coefficients are given up to  ${\cal
O}(\epsilon)$.

\subsection{gluon-gluon scattering}

For the interference of tree with one-loop for gluon-gluon scattering we  find,
\bea
B^{gg}_{11} &=& \frac{2}{\eps}C_A\nn \\
B^{gg}_{10} &=& C_A \left(-{4\over \eps^2} +\left(-{67\over 9} + \pi^2 \right)
+ \left(-{422\over 27} + 2 \zeta_3\right) \eps + \left(-{2626\over 81} +
{\pi^4\over 15}\right) \eps^2\right) \nn \\
&+& N_F \left({10\over 9} + {74 \over 27} \eps + {580\over 81} \eps^2\right) + \beta_0 \left(-{2\over \eps}\right)\nn \\
D^{gg}_{10} &=& B^{gg}_{11}
\eea
while the two-loop coefficients are given by,
\bea
B^{gg}_{22} &=& {1\over 2} \left( B^{gg}_{11}\right)^2\nn \\
B^{gg}_{21} &=& B^{gg}_{11} B^{gg}_{10} + C_A \beta_0 {2\over \eps^2} + C_A K {2
\over \eps}
+ C_A^2 \left({404\over 27} - 2\zeta_3\right) + C_AN_F \left(-{56\over 27}\right) \nn \\
B^{gg}_{20} &=& {1\over 2} \left( B^{gg}_{10}\right)^2 - C_A\beta_0 {2\over
\eps^3} - \beta_0^2 {2\over \eps^2}\nn\\
&+& C_A^2 \left(\left(-{67\over 9} -{38\pi^2\over 3}\right){1\over
\eps^2} + \left(-{1276\over 27} +2\zeta_3 + {44\pi^2\over 9}\right) {1\over \eps}
\right.\nn \\
&& \left.
+ \left(-{12433\over 81} + {44} \zeta_3 + {268\pi^2\over 27} - {19\pi^4
\over 45}\right)\right) \nn \\
&+& C_AC_F \left({24\pi^2\over \eps^2}\right) \nn \\
&+& C_AN_F \left( {10\over 9\eps^2} + \left({109\over 9}-{8\pi^2\over 9} \right)
{1\over \eps} + \left( {3169\over 81}  - {40\pi^2 \over 27}
\right)\right) \nn \\
&+& C_FN_F \left({1\over \eps} + \left({55 \over 6} - 8 \zeta_3 \right)\right) \nn \\
&+& N_F^2 \left(-{20 \over 27\eps}-{22\over 9}\right)  \nn \\
D^{gg}_{21} &=& \left(B^{gg}_{11}\right)^2 \nn \\
D^{gg}_{20} &=& B^{gg}_{21}
\eea

\subsection{quark-gluon scattering}

For the interference of tree with one-loop for quark-gluon scattering we  find,
\bea
B^{qg}_{11} &=& \frac{2}{\eps}C_A\nn \\
B^{qg}_{10} &=& C_A \left(-{2\over \eps^2} + 1 + \pi^2 + 
\left({5\over 3} + 2\zeta_3\right)\eps + \left({25\over 9} + {\pi^4\over 15}
\right) \eps^2 \right) \nn \\
&+& C_F \left(-{2\over \eps^2} - {3\over \eps} -8 -16 \eps -32 \eps^2\right) \nn \\
&+& N_F \left({1\over 3} \eps + {14\over 9} \eps^2\right) \nn \\
D^{qg}_{10} &=& B^{qg}_{11}
\eea
while the two-loop coefficients are given by,
\bea
B^{qg}_{22} &=& {1\over 2} \left(B^{qg}_{11}\right)^2\nn \\
B^{qg}_{21} &=& B^{qg}_{11}B^{qg}_{10} + C_A \beta_0 {2\over \eps^2} + C_A K
{2\over \eps}+ C_A^2 \left({404\over 27} - 2 \zeta_3 \right) + C_AN_F \left(-{56\over 27} \right) \nn \\
B^{qg}_{20} &=& {1\over 2} \left( B^{qg}_{10}\right)^2 -C_A \beta_0 {1\over
\eps^3} - C_F \beta_0 {1\over \eps^3} + \nn \\
&+& C_A^2 \left(\left( -{67\over 18} - {17\pi^2\over 6} \right) {1\over \eps^2}
+ \left( -{94\over 27} + {77\pi^2\over 18} + \zeta_3\right) {1\over \eps}+ \left( {1289\over 162} + {469\pi^2\over 54} + {34\zeta_3\over 3} - {29\pi^4
\over 60}\right)\right) \nn \\
&+& C_AC_F \left(\left( -{83\over 9} + {25\pi^2\over 6} \right) {1\over \eps^2}
+ \left( -{4129\over 108} -{11\pi^2\over 18} + 13 \zeta_3 \right) {1\over \eps}
\right.\nn \\
&&\left.
+ \left(-{91765\over 648} -{110\pi^2\over 27} + {184\zeta_3\over 3} + {11\pi^4
\over 36} \right)\right) \nn \\
&+& C_F^2 \left(\left(-{3\over 4} +\pi^2 -12 \zeta_3 \right) {1\over \eps} +
\left( -{1\over 8} + {29\pi^2\over 6} -30\zeta_3 -{11\pi^4\over 45} \right)\right) \nn \\
&+& C_AN_F \left({5\over 9\eps^2} + \left( {1\over 27} - {7\pi^2\over 9} \right)
{1\over \eps} + \left( -{178\over 81} -{35\pi^2\over 27} - {16\zeta_3\over 3}
\right)\right) \nn \\
&+& C_FN_F \left({14\over 9\eps^2} + \left( {353\over 54} + {\pi^2\over 9}
\right) {1\over \eps} + \left( {7541\over 324} +{14\pi^2\over 27} -
{4\zeta_3\over 3}\right)\right) \nn \\
&+& N_F^2 \left(-{2\over 9}\right)  \nn \\
D^{qg}_{21} &=& \left(B^{qg}_{11}\right)^2 \nn \\
D^{qg}_{20} &=& B^{qg}_{21}
\eea

\subsection{quark-quark scattering}

For the interference of tree with one-loop for quark-quark scattering we  find,
\bea
B^{qq}_{11} &=& \frac{2}{\eps}C_A\nn \\
B^{qq}_{10} &=& C_A \left(\left({85\over 9} + \pi^2\right) +\left({512\over 27}
+ 2 \zeta_3 \right) \eps + \left( {\pi^4\over 15} + {3076 \over 81}
\right)\eps^2\right) \nn \\
&+& C_F \left(-{4\over \eps^2}-{6\over \eps} - 16 - {32\eps} -{64\eps^2} \right) \nn \\
&+& N_F \left(-{10\over 9} - {56\eps\over 27} - {328 \eps^2 \over 81}\right) \nn \\
&+&  \beta_0 \left({2\over \eps} \right)\nn \\
D^{qq}_{10} &=&  -{4\over \eps}C_A+{16 \over \eps} C_F 
\eea
while the two-loop coefficients are given by,
\bea
B^{qq}_{22} &=& \frac{1}{2} \left(B^{qq}_{11}\right)^2\nn \\
B^{qq}_{21} &=& B^{qq}_{11}B^{qq}_{10}+ C_A\beta_0 {2\over \eps^2} +C_A K
{2\over \eps} + C_A^2 \left({404\over 27} - 2 \zeta_3 \right) + C_AN_F \left(-{56 \over 27}\right) \nn \\
B^{qq}_{20} &=& \frac{1}{2} \left( B^{qq}_{10}\right)^2 - C_F\beta_0 {2\over
\eps^3} + \beta_0^2 {2\over \eps^2}\nn\\
&+& C_A^2 \left(-{2\pi^2 \over \eps^2} +\left( {1088\over 27} + {11\pi^2\over
3}\right){1\over \eps} + \left( {4574\over 27} + {67\pi^2\over 9} -{64
\zeta_3\over 3}
-{49\pi^4\over 90}\right)\right)  \nn \\
&+& C_AC_F \left(\left( -{166\over 9} + {37\pi^2\over 3} \right) {1\over \eps^2}
+ \left( -{4129\over 54} -{11\pi^2\over 9} + 26\zeta_3\right) {1\over \eps}
\right. \nn \\
&& \left. + \left(-{91765\over 324} -{220\pi^2\over 27} + {368\zeta_3\over 3} + {11\pi^4
\over 18} \right)
\right) \nn \\
&+& C_F^2 \left(-{24\pi^2\over \eps^2} + \left( -{3\over 2} + 2\pi^2 -24\zeta_3
\right){1\over \eps} + \left( -{1\over 4} + {29\pi^2\over 3} -60\zeta_3
-{22\pi^4\over 45} \right)\right) \nn \\
&+& C_AN_F \left(\left( -{325\over 27} -{2\pi^2\over 3} \right) {1\over \eps} 
+\left( -{1175\over 27} - {10\pi^2\over 9} - {32\zeta_3\over 3}\right)\right) \nn \\
&+& C_FN_F \left({28\over 9\eps^2} + \left({326\over 27} + {2\pi^2\over
9}\right){1\over \eps}\right) +\left({3028\over 81} + {28\pi^2\over 27} +
{16\zeta_3\over 3} \right)\nn \\
&+& N_F^2 \left({20\over 27\eps} + 2\right)  \nn \\
D^{qq}_{21} &=&  -{8\over \eps^2}C_A^2 + {32 \over \eps^2}C_AC_F \nn \\
D^{qq}_{20} &=&  {-2(C_A-4C_F)\over C_A} B^{qq}_{21}
\eea

\end{document}